\documentclass{aa}

\usepackage{graphicx}
\usepackage{txfonts}
\usepackage{amsmath}	% Advanced maths commands
\usepackage{amssymb}	% Extra maths symbols
\usepackage{newtxtext,newtxmath}

\usepackage[shortlabels]{enumitem}

\newcommand{\MG}{R-MOND}
\newcommand{\muval}{\mathrm{log}_{10}[\mu_{(1/\mathrm{kpc})}]}

\newcommand{\GN}{\ensuremath{G_{{\rm N}}}}
\newcommand{\Phih}{\tilde{\Phi}}
\newcommand{\grad}{\ensuremath{\vec{\nabla}}}
\newcommand{\Gt}{\ensuremath{\tilde{G}}}

\newcommand{\Fcal}{{\cal F}}
\newcommand{\Jcal}{{\cal J}}

\newcommand{\Qcal}{{\cal Q}}
\newcommand{\Ycal}{{\cal Y}}
\newcommand{\KB}{\ensuremath{K_{{\rm B}}}}

\usepackage[colorinlistoftodos]{todonotes}  % For commenting on the pdf.
\usepackage{xcolor}

\usepackage{lineno}
\usepackage{cuted}  % For long equations

\usepackage{setspace}

\usepackage{float}  % for positioning figures
\usepackage{ulem}  % for canceling out text.
\def\bi{\begin{itemize}}
\def\ei{\end{itemize}}

\begin{document}

    \titlerunning{Orbits in R-MOND}
   \title{Extension of General Relativity with MOND limit predicts novel orbital structure in and around galaxies}

   \author{C. Llinares
          \inst{1}
          }

   \institute{Departamento de Física, Faculdade de Ciências, Universidade de Lisboa, Edifício C8, Campo Grande, PT1749-016 Lisbon, Portugal \\
              \email{cllinares@fc.ul.pt}
        }

   \date{Received XXX; accepted YYY}

% \abstract{}{}{}{}{}
% 5 {} token are mandatory
   \abstract
  % context heading (optional)
  % {} leave it empty if necessary  
   {Detailed knowledge of the different classes of stellar orbits that can be accommodated in a given galactic potential is a prerequisite when building self-consistent models using for instance the Schwazschild technique.  Furthermore, observational properties of galaxies depend on what these classes of orbits are and on the presence of chaos in the systems.  In the realistic case in which the starting point for modeling is not a gravitational potential, but an observed density distribution, we will require a gravitational theory to make the connection between the stars that we see and the movement these stars may be having.  The argument can be turned upside down:  understanding what orbits may be allowed by each gravitational theory may give us a greater insight on what these theories are and on how we can test them.}
  % aims heading (mandatory)
   {Our aim is to understand novel properties of orbits that are predicted by the latest extension of the MOND phenomenology into the relativistic world.}
  % methods heading (mandatory)
   {We integrated orbits numerically in a fixed density distribution.  The potential required for such integration was obtained also numerically by assuming different gravitational models.}
  % results heading (mandatory)
   { Thanks to the presence of a mass term in the field equations, the theory can allocate new classes of orbits that do not exist in Newtonian gravity nor standard MOND.  We discuss consequences that these new families of orbits can have in non-linear cosmological structure formation as well as explore a possible alternative model for galactic structure based on them.}
  % conclusions heading (optional), leave it empty if necessary
   {}

   \keywords{Gravitation -- 
   Galaxies: kinematics and dynamics -- 
   Galaxies: structure -- 
   Cosmology: dark matter -- 
   Cosmology: large-scale structure of Universe --
   Chaos}
   
   \maketitle

\section{Introduction}

Broadly speaking, the scientific method is based on the idea that given a \textit{theory}, it is possible to determine \textit{predictions}, which can be compared with \textit{data} whenever they become available (ideally, after these predictions were made).  Discrepancies between predictions and data may be related to three independent issues:  there may be a problem with the data (e.~g.~a source of error that was not taken into account when observing), a problem with the predictions (e.~g.~a dominant effect not taken into account when doing the calculations) or a problem with the theory itself.  In the context of gravity and the dynamics of astrophysical systems, discrepancies between predictions and data were found at the beginning of 20th century \citep{1933AcHPh...6..110Z, 1937ApJ....86..217Z, 2017arXiv171101693A}.  After few decades of public debate \citep{2017NatAs...1E..59D}, consensus was reached and it was accepted that the discrepancy did not originate in the data, but in the predictions:  a source of energy was not taken into account when calculating the relation between the amount of matter that was present in the systems and the magnitude of the gravitational force.

The story around the concept of dark matter is very compelling and, in fact, gave rise to a successful cosmological model years later.  However, the fact that the \textit{only} evidence that we have for the presence of this additional component in the Universe is through gravity may set doubts on its existence as a real entity.  We had to wait until the last stages on the 20th century for someone to explore the possibility that the problem may not be with the predictions, nor the data, but with the theory that was under scrutiny \citep{Milgrom83, 1984ApJ...286....7B}.  

Milgrom's gravitational theory, dubbed Modified Newtonian Dynamics, was proposed as a fitting formula for rotation curves of galaxies:  when accelerations are below a fixed threshold, the theory predicts a logarithmic gravitational potential that produces flat rotation curves in galaxies (as observed).  This phenomenology can be represented with a single non-linear equation, but very quickly MOND became a family of gravitational theories that consists in a multiplicity of relativistic and non-relativistic Lagrangians  \citep[e.g.][]{1984ApJ...286....7B, 1997ApJ...480..492S, 2007PhRvD..75d4017Z, Bekenstein_2004-PhRvD_70h3509B, 2015JCAP...12..026B, 2011PhRvD..84l4054D, 2009PhRvD..80l3536M, 2009ApJ...698.1630M, 2015PhRvD..91b4022K, 2017ScPP....2...16V}.  
While these Lagrangians can provide strong theoretical grounds to the MOND idea, they have difficulties in providing a fully working cosmology.  For this reason, the astronomical community largely focused on the implications of MOND in galaxies and in testing the framework mainly at these small scales.

From the point of view of cosmology, the apparent stagnation in theoretical developments have recently moved forward: \cite{Skordis_Zlosnik_2021} have finally managed to write a Lagrangian that can do background cosmology and provide predictions that are consistent with CMB and matter power spectrum observations.  Furthermore, the Lagrangian has been found to be free of instabilities \citep{Skordis:2021vuk}.  Having proven that the theory is viable at a linear level, it is time to understand its consequences in both, individual objects such as galaxies as well as non-linear cosmology.

On the galaxy side, the study of the families of stellar orbits that can be accommodated in this model is crucial for several reasons.  First, Schwarzschild techniques for constructing dynamical models of galaxies \citep{1979ApJ...232..236S} require the determination of libraries of orbits, which can be built only after understanding the possibilities of the theory.  Also, the effects of chaos on the overall properties of the systems depends on these families of orbits.  Detailed analysis of the orbital structure of galaxies can be found for instance in works by \cite{1999AJ....118.1177M, 1998ApJ...498..625M, 2007CeMDA..99..307A, 2005CeMDA..91..173M, 2004CeMDA..88..379M, 1998MNRAS.298....1C}.  Orbits in disk galaxies has been studied by \cite{2013CeMDA.116..417Z, 2022MNRAS.509.1465P}.  Furthermore, the impact of chaos in these classifications was studied by \cite{1996ApJ...471...82M, 1996Sci...271..337M, 2009ASSP....8..203M, 1999CeMDA..73..159C, 2003CeMDA..85..247C, 2014MNRAS.438.2871C, 2020MNRAS.492.4398M, 2020MNRAS.495.1608C}.  Non-specialists in stellar orbits may find additional information in textbooks by \cite{Binney_Tremaine_2008} or \cite{2021isd..book.....C}.

When it comes to study orbits in the original MOND model, it is important to take into account that the MOND gravitational potential associated to a galaxy is expected to mimic that of the galaxy plus a dark matter halo (when interpreted in the context of Newtonian gravity).  Since the potential is approximately the same in both gravitational theories, the stellar orbits are expected to be the same apart from minor perturbations that are a result of the exact definition that we employ for MOND.  In the case of the version of MOND that we will study here, things are different.  The theory was designed so that it can reproduce the MOND phenomenology, however, the equations are not exactly as in MOND.  In particular, they contain a mass term which provides the dynamical fields with spacial oscillations.  These oscillations are translated into oscillations in the force profile, which is ultimately the field that defines the trajectory of free particles (i.e. orbits).  Here we are interested in understanding the new classes of orbits that arise from these spacial oscillations. We will achieve this by fixing a baryonic distribution to represent an elliptical galaxy and will study orbits in the resulting potentials in increasing levels of complexity, from 1D spherical potentials to 3D perturbed disks.

Before we close this introduction, we would like to make a clarification of language.  Strictly speaking, the theory that we will study in this work \citep[defined by][]{Skordis_Zlosnik_2021} is not MOND (MOND is a phenomenological non-relativistic framework, while here we are dealing with a modification of general relativity which has a non-relativistic MONDian limit). Thus, through this paper, we will refer to this theory as Relativistic MOND (R-MOND).  The paper is structured as follows: in Section \ref{section:models} we describe the relevant details of {\MG} and the baryonic distribution that will be used for the numerical experiments.  Section \ref{section:results} shows our catalog of orbits, where we focus in particular on new orbits that exist only in this model.  Section \ref{section:discussion} presents a discussion on the implications of these new orbits in the context of non-linear cosmological structure formation, the orbital structure of disk galaxies, and observational consequences.  We finally present our conclusions in Section \ref{section:conclusions}.

\section{Gravitational and galactic models}
\label{section:models}

We are interested in studying stellar orbits in a gravitational theory defined by the following Lagrangian \citep{Skordis_Zlosnik_2021}:

\begin{align}
    S =&  \int d^4x \frac{\sqrt{-g}}{16\pi \Gt}\bigg\{ R  - 2 \Lambda
     - \frac{\KB}{2}  F^{\mu\nu} F_{\mu\nu}
    + 2  (2-\KB) J^{\mu} \nabla_\mu \phi 
    \nonumber
    \\
    &
    - (2-\KB) \Ycal
    - \Fcal(\Ycal,\Qcal)
     - \lambda(A^\mu A_\mu+1)\bigg\}  + S_m[g], 
        \label{eq:action}
    \end{align}
where $J^\mu \equiv A^\nu \nabla_\nu A^\mu$ and $F_{\mu\nu} \equiv 2\nabla_{[\mu} A_{\nu]}$, $g$ is the metric determinant, $\nabla_\mu$ the covariant derivative compatible with $g_{\mu\nu}$,   $R$  is the Ricci scalar, $\Lambda$ is the cosmological constant,
$\Gt$ is the bare gravitational strength, $\KB$ is a constant denoting the  vector field coupling strength and $\lambda$ is a Lagrange multiplier imposing the unit time-like constraint on $A_{\mu}$.
The matter action $S_m$ is  assumed not to depend explicitly on $\phi$ or $A^\mu$.  Furthermore, two additional scalars are defined:  $\Qcal = A^\mu \nabla_\mu \phi$ and 
$\Ycal= (g^{\mu\nu} + A^{\mu} A^{\nu})\nabla_\mu \phi \nabla_\nu \phi$.  Finally, the free function $\Fcal$ can be constrained by imposing appropriate limits on the fields.  Note that the theory has similarities with the very well known TeVeS model \citep{Sanders_1997-ApJ480,Bekenstein_2004-PhRvD_70h3509B}, in particular in the number of fundamental fields employed (i.e. the vector field $A^{\mu}$ and the scalar $\phi$ in addition to the metric $g_{\mu\nu}$).  The stability of the model has been discussed in \cite{Skordis:2021vuk}.

\cite{Skordis_Zlosnik_2021} have shown that in the weak field limit, the equation for the metric perturbations $\Phi$ that are responsible for the dynamics of galaxies (i.e. the Poisson's equation for what we call gravitational potential in the context of galaxy dynamics) is substituted with the following system
\begin{align}
    \label{eq_1_mond}
    \grad^2\Phih +   \mu^2  (\tilde{\Phi} + \chi) =& \frac{8\pi \Gt}{2-\KB} \rho_b,\\
    \label{eq_2_mond}
    \grad \cdot \left[f\left(\frac{\grad\chi}{a_0}\right)\grad \chi\right] +  \mu^2  (\tilde{\Phi} + \chi) =& \frac{8\pi \Gt}{2-\KB} \rho_b, 
\end{align}
where $f = \frac{d\Jcal}{d\Ycal}$ and we introduced for convenience a new potential $\chi$ which combines derivatives of $\phi$ and components of $A^{\mu}$.  Furthermore, we defined an auxiliary field $\tilde{\Phi} \equiv \Phi - \chi$ (see \cite{Skordis_Zlosnik_2021} or \cite{Verwayen} for details on the derivation of these equations).
In the spherical coordinates in which we will solve the field equations, the system takes the following form:
\begin{align}
    \frac{d^2\tilde{\Phi}}{dr^2} &= \frac{4 \pi \GN}{1+\beta_0} \rho_b - \mu^2(\tilde{\Phi} + \chi) - \frac{2}{r}\frac{d \tilde{\Phi}}{dr}
\label{d2_Phi_spher_sym}
\\
\left(x\frac{df}{dx}+f(x) \right)    \frac{d^2\chi}{dr^2} &= \frac{4 \pi \GN}{1+\beta_0} \rho_b - \mu^2(\tilde{\Phi} + \chi) - \frac{2}{r}f(x)\frac{d\chi}{dr}, 
\label{d2_chi_spher_sym}
\end{align}
where the last term in the right-hand side is the part of the differential operator that arises from the change of coordinates.  

These equations provide a prescription for dark matter effects, but differ from the original MOND equation \citep{Milgrom83, 1984ApJ...286....7B} not only in the exact definition they give for these effects, but more fundamentally thanks to the presence of a mass term $\mu^2(\tilde{\Phi}+\chi)$.  The aim of this paper is to study the consequences of this term in the behavior of particles in and around galaxies.  To this end, we will integrate the Hamilton equations associated to the total force:
\begin{align}
\dot{\textbf{x}} & = \mathbf{p} \\
\dot{\textbf{p}} & = \mathbf{F}(x,y,z) = \nabla\Phi, 
\end{align}
where $x$, $y$ and $z$ are the Cartesian components of the position vector $\textbf{x}$.  Since we are interested in describing only new classes of orbits that can be accommodated in these potentials, 
we neglect additional cosmological expansion factors that will appear in both the field and Hamilton equations in the more general case.  Including these terms will slightly change the exact numerical results, but will not change the orbital structure.

Throughout the paper, we assume ${\muval}=-1.5$ except where we mention a different value.  \cite{Verwayen} have shown that changing the value of $\mu$ changes the radius at which the mass term becomes dominant and activates spacial oscillations in the potentials and force: larger values of $\mu$ imply smaller transition radius towards the oscillatory regime.  A priori one may expect that this dependence could be eliminated by rescaling the distances.  However, this is not possible thanks to the presence of additional scales in both the density profile and the theory itself (which contains the original acceleration scale $a_0$ present in the classical MOND Lagrangians).  All this means that the orbital structure will necessarily have a dependence on $\mu$.  However, such variation is expected to be in the exact numbers (for instance in the ratio of chaotic to regular orbits) and not in the overall phenomenology of the orbits that we will describe in following sections.

The standard MOND equation has analytic solutions for the spherical case, however, the presence of the mass term forced us to calculate numerical solutions.  To this end, we used the Runge-Kutta solver order 8 provided by the \texttt{scipy} library as a wrapper for ODEPACK \citep{hindmarsh1982odepack}. The same routines were employed to solve the Hamilton equations.  The force required was obtained with a cubic interpolation scheme from the same library.  The Hamilton equations are known to have chaotic solutions which may be real or the consequence of an unstable numerical scheme.  In order to confirm that our solutions are stable from a numerical point of view, we compared our results for several examples with a second order leap-frog scheme with a fixed time step written in house.

In order to solve these equations, we need to fix the density profile, for which we chose a Hernquist model  \citep{Hernquist_1990}:
\begin{equation}
\rho_b(r) = \frac{8 \rho_b(a_H)}{\frac{r}{a_H}\left(1+\frac{r}{a_H}\right)^3}.  
\end{equation}
See \cite{Binney_Tremaine_2008} for details on the Hernquist profile and note that our definition of the normalization of the density contains a factor of 8 that comes from the fact that we define the normalization as the density evaluated at $r=a_H$.

For the free parameters, we assume values provided by \cite{2005MNRAS.362..197T} for the luminosity density and angular scale of the elliptical galaxy NGC 1379: $(\log_{10}(\tilde{\rho}(a_H)[L{\sun}/\mathrm{pc}^3]), \tilde{a}_h) = (-1.03, 18.5 ~ \mathrm{arcsec})$.  These values can be translated into our mass density profile by taking into account the distance modulus provided by the same authors: 31.51.  The final values that we use for our calculations are thus $(\log_{10}(\rho_b(a_H)[M_{\odot}/\mathrm{kpc}^3]), a_H) \sim (7.79, 1.8 ~ \mathrm{kpc})$, where we assumed a mass-to-light ratio equal to 1 $M_{\odot}/L_{\odot}$.  Since the main interest of this paper are solutions obtained with the MOND family of gravitational theories, we do not take into account a dark matter profile associated to this baryonic profile (these effects will be instead provided by the Lagrangian that describes gravity).  Only in the case where we show Newtonian solutions, the presence of a dark matter halo will change the exact values associated with our solutions, but will not change the phenomenology that will be discussed.

We will study orbits in this spherical potential as well as flattened and triaxial potentials derived by evaluating the spherical solution for the potential in a modified radius $k$ defined as:
\begin{equation}
k(x,y,z) = \sqrt{x^2 + \frac{y^2}{b^2} + \frac{z^2}{c^2}}. 
\label{def_k} 
\end{equation}
Thus, the force in the most general triaxial case is given by:
\begin{equation}
\mathbf{F}(x,y,z) = \left( 1, \frac{1}{b}, \frac{1}{c} \right) \times \frac{1}{r} \frac{d\Phi}{dr}.
\label{def_flattened_force}
\end{equation}

\begin{figure}[!t]
    % REFEREE
    %\includegraphics[width=0.8\columnwidth]{orbits.pdf}
    \includegraphics[width=\columnwidth]{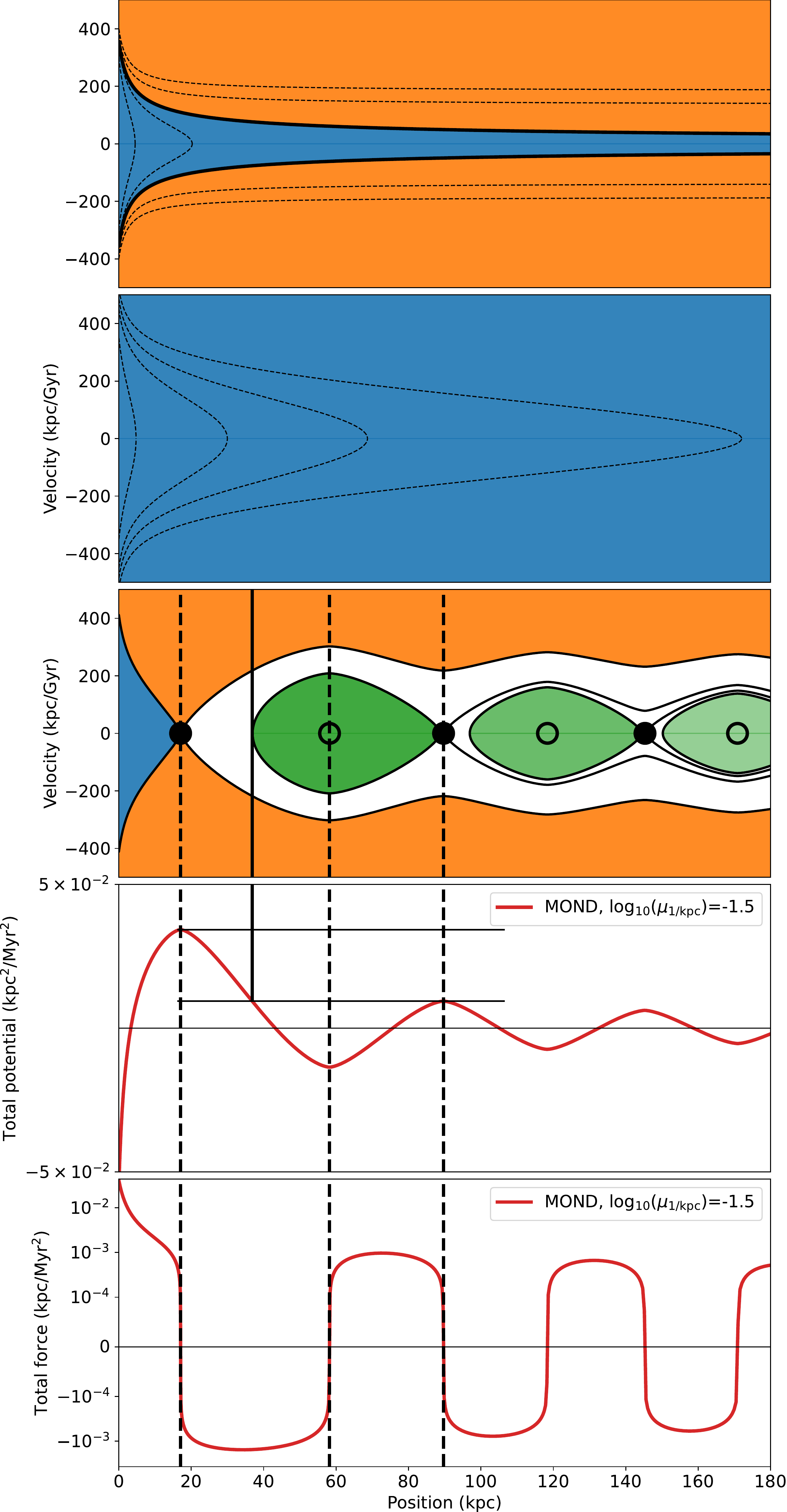}
    \caption{The first three panels show the orbital structure in phase-space for radial orbits in a 1D spherical system; from top to bottom, these panels correspond to Newtonian gravity, {\MG} with $\mu=0$ and {\MG} with $\muval = -1.5$.  The thick lines are the orbits that separate different regions in phase-space.  The colors of these regions are as follows: blue~=~bound orbits, orange~=~free particles, white~=~expelled particles, green~=~perturbed circular orbits.  The thin dotted lines are additional orbits given as example.  The black dots in the third panel are stable (open) and unstable (filled) equilibrium points where the force is equal to zero.  The two bottom panels show the total potential $\Phi$ and force profile for the {\MG} model with $\muval = -1.5$.  The dotted vertical lines are the zeros of the force profile that define the equilibrium points, while the continuous lines show how to determine the transition between empty and green regions in the third panel.} 
    \label{fig:orbits}
\end{figure}

\section{Results: a catalog of orbits in {\MG}}
\label{section:results}

\subsection{1D orbits in spherical systems}
\label{section:orbits}

The simplest orbits one can think of when building a galaxy are 1D radial orbits, so we start our discussion from there.  The three upper panels of Figure \ref{fig:orbits} show a decomposition of the 2D phase-space $(r, v_r)$ according to the type of orbits that can be allocated in the galactic potential described in Section \ref{section:models}.  Each color panel corresponds to a different gravitational theory.  From top to bottom we have: Newtonian gravity, {\MG} with $\mu=0$ (representing classical MOND), and {\MG} with $\muval = -1.5$.  We use the following coloring scheme for each individual region: blue~=~bound orbits, orange~=~free particles, empty~=~expelled particles, green~=~perturbed circular orbits.

In the Newtonian case (upper color panel), we have the usual decomposition of the phase space for spherical objects.  Orbits with a velocity smaller that the escape velocity are bound (blue region) and the remainder are free particles (orange region).  The thick black line corresponds to the central escape velocity, which is given by:
\begin{equation}
v_e = 16 a_H \sqrt{\pi G \rho_b(a_H)} \sim 355 \mathrm{km}/\mathrm{sec}.   
\end{equation}
In order to calculate this value, we took into account that the Newtonian gravitational potential is given by
\begin{equation}
\Phi(r) = -\frac{16\pi G a_H^2 \rho_b(a_H)}{1+r/a_H}.
\end{equation}
Additional bound and unbound orbits are shown as thin dashed lines as further examples.

The second panel from the top shows orbits in the {\MG} case with $\mu=0$ (which corresponds to the classical MOND limit of {\MG}).  Since the potential becomes logarithmic far from the galaxy, there is no escape velocity and thus, all orbits are bound.  This means that particles will end up approaching the galaxy no matter how fast they may be moving away from it.  Cosmological simulations show that the effect does have an impact on the overall cosmological evolution, resulting in redshift zero boxes that contain almost not structures resembling a cosmic web and containing instead only a few very large and massive objects \citep{Nusser02, llinares_thesis, 2016MNRAS.460.2571C}.

The third panel of the same figure shows the orbital structure of the {\MG} theory when assuming $\muval=-1.5$.  Before discussing these orbits, we describe the potential and force profiles responsible for them, which are shown in the two bottom panels of Figure \ref{fig:orbits}. The inner part of the profile is as in standard gravity and classical MOND (i.e. a potential well is associated to the galaxy which can keep it as an integral entity).  Farther away from the center, when the derivatives of the potential and mass density are small enough, the mass term in the field equations dominates and thus, the fields oscillate around zero.  The fact that the force profile becomes vertical when crossing zero is not related to the symlog scaling that we used for the vertical axis, but is actually physical and related to the fact that the factor between parenthesis in the left-hand side of Eq.~\ref{d2_chi_spher_sym} is equal to zero at these points (this forces the second derivative of the potential to be infinite at these specific points).  More details on these profiles as well as their dependence on the model parameters are described by \cite{Verwayen}.

The structure that the {\MG} force distribution induces in the phase space of radial orbits is shown in the third panel of Figure \ref{fig:orbits}.  The first difference with respect to the other two panels is the presence of equilibrium points that arise from the oscillations in the potentials and where the gravitational force is equal to zero.  The empty and filled circles in the horizontal axis of this panel correspond to elliptic and hyperbolic points respectively (i.e. stable and unstable equilibrium points).  An interesting fact related to the hyperbolic points is that the unstable direction that moves away from them to the left goes around the corresponding green region and comes back to the same point from the stable direction.  This means that these directions are unstable at first order, but stable when the complete orbit is taken into account.  That is not the case for the unstable directions on the right-hand side of these points;  in these cases, unstable orbits are effectively unstable up to infinity.  Particles located in empty circles will be in stable equilibrium and thus, will not move from these points in the radial direction.  Infinitesimal tangential velocity will result in circular orbits (we will study this in more detail in the following section).

Regarding the decomposition of phase-space, here we go from the two or one region that we had in the Newtonian or classical MOND cases to four distinct regions.  The most important difference between classical MOND and this case is that the size of the region where particles are bound to the galaxy (blue region to the left) is now comparable to the bulk of the bound region that we find in the Newtonian case.  Furthermore, we recover a region of phase-space where particles are free and thus, can cross the galaxy, but not necessarily remain bound to it.  Moreover, two additional regions can be found in this panel.  In the empty regions to the right of the first hyperbolic point, the force oscillates in space as we move outwards, but eventually behaves as a net repulsive force which expels particles from the surroundings of the galaxy.  Incoming particles in this particular region will approach the galaxy until a minimum distance and then will be expelled in the direction they came from without having a chance to enter the galaxy (we provide more details on these kinds of orbits in Section \ref{section:incoming}).

Finally, in the green regions, particles oscillate around the stable equilibrium points with orbits that are closed in the plane $r-v_r$.  In the next section we will see that providing particles with a tangential velocity will result in perturbed circular orbits whenever this velocity is below the escape velocity.  These two regions (empty and green) are arranged in a fractal like structure that repeats itself towards the right.  We identify each substructure with a different intensity for the colors.

\begin{figure*}
    \begin{center}
    \includegraphics[width=\textwidth]{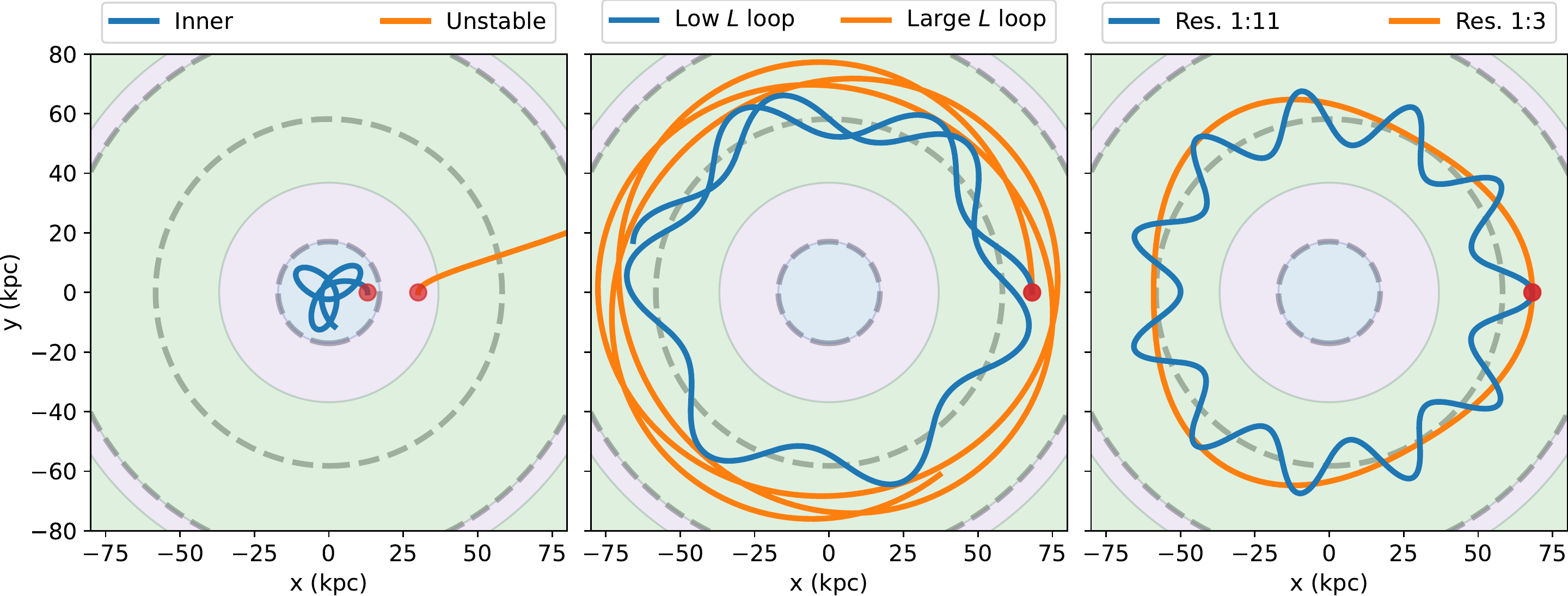}
    \caption{Classification of 2D orbits in a spherical system.  The gray dashed circles correspond to the zeros of the force profile and divide the configuration space in regions with attractive and repulsive gravitational force.  The colors in the background distinguish the same regions defined in the third panel of Figure \ref{fig:orbits} for the same model and with the same value for the mass parameter. The red dots represent the initial conditions that was chosen for each orbit.}
    \label{fig:spherical_2d}
    \end{center}
\end{figure*}

\begin{figure}
    \begin{center}
    \includegraphics[width=\columnwidth]{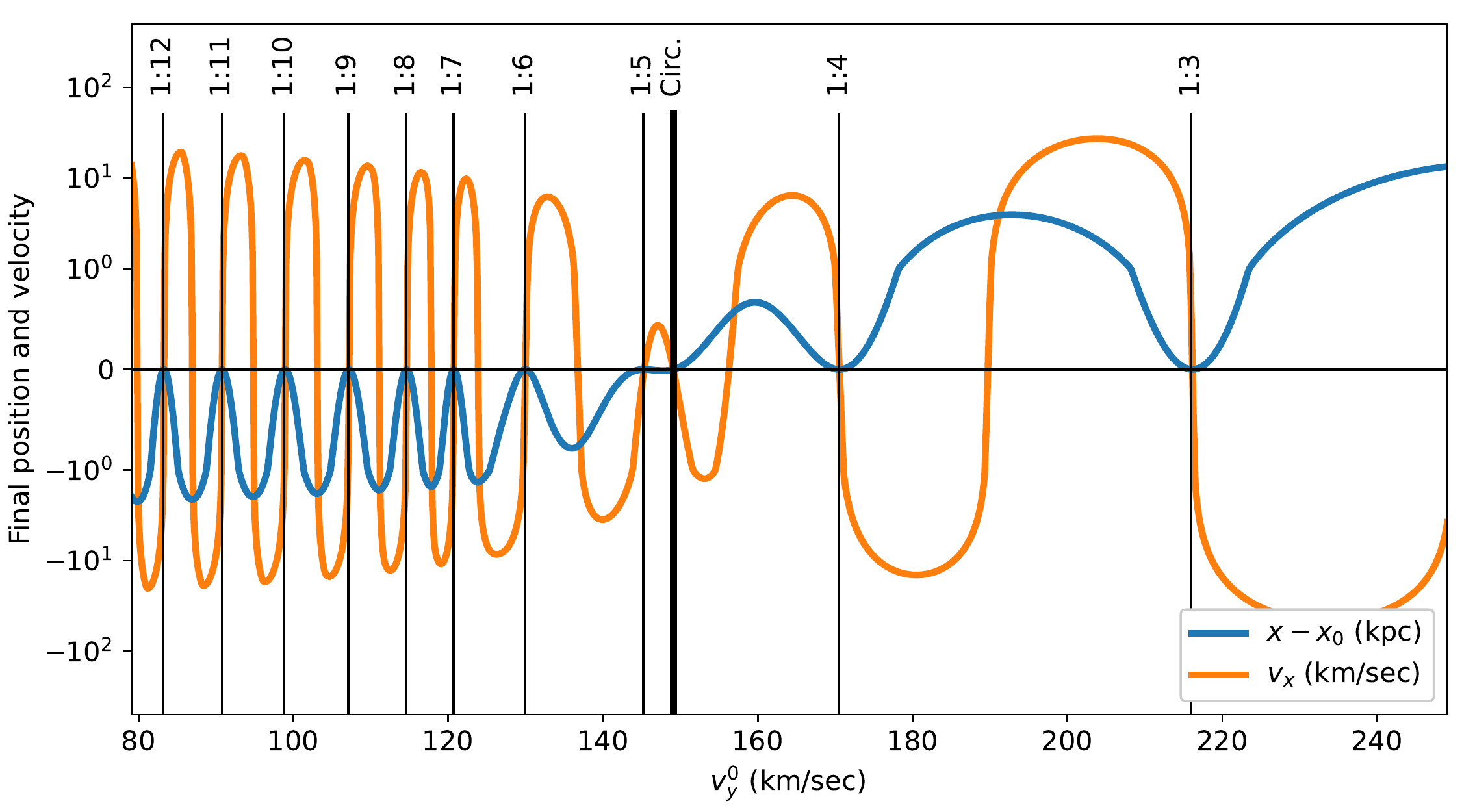}
    \caption{Distribution of $1:n$ resonances in velocity space for a spherical system.  The initial conditions in configuration space are the same for each velocity and are located in the positive $x$ axis.  The blue line is the deviation for this initial condition after an entire orbital period;  the orange line is the final velocity in the tangential direction.  The initial velocities for which both curves are zero correspond to resonances and are highlighted with vertical lines.  The thick vertical line is the circular velocity.  We used a symlog scaling for the vertical axis which switches to a linear scaling at $\pm 1$.}
    \label{fig:spherical_2d_resonances}
    \end{center}
\end{figure}

\begin{figure*}
    \includegraphics[width=\textwidth]{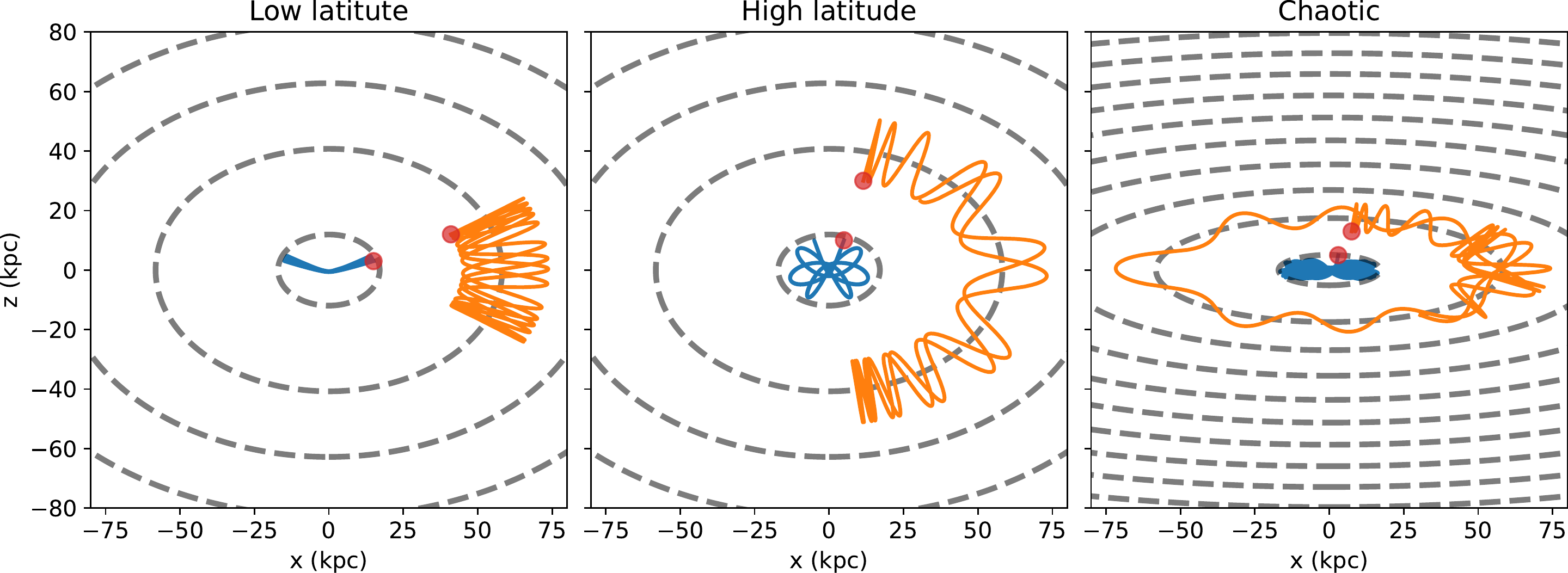}
    \caption{Examples of box orbits in a flattened 2D system.  The ellipsoid is oriented with its minor axis in the $z$ direction.  We show orbits in the $x-z$ plane.  The blue and orange orbits belong to the inner region (i.e. before the oscillations in the force profile start) and the first stable ring respectively.  The dashed lines are the zeros of the force.}
    \label{fig:flattened_2d_box}
\end{figure*}

The existence of the unstable empty regions may be surprising given that particles are simply oscillating in a series of potential wells, so one would expect the green regions to be interconnected as it happens for instance in the phase space of a pendulum.  However, we need to take into account that the height of the peaks in the potential depends on which peak we are looking at (see horizontal lines highlighting two of these peaks in the fourth panel of Figure \ref{fig:orbits}).  Thus, it is possible to have particles that have zero velocity, but enough potential energy to jump the potential well to the right. These are the particles that will eventually populate the empty regions.

The continuous horizontal and vertical line in the three bottom panels in Figure \ref{fig:orbits} show that the transition between the unstable and stable regions (empty/green) can be determined by taking into account the excess energy of particles located in the unstable region. Particles between the first maxima of the potential and the continuous vertical line have enough energy to jump out of the potential well to the right and escape to infinity.  The other three vertical dashed lines in these panels show the position of three equilibrium points and their relation with maxima/minima of the total potential and the zeros of the force profile.

\subsection{2D orbits in spherical potentials}

A first extension of the orbits presented in the previous section can be obtained by keeping the spherical symmetry and providing the particles with a tangential velocity in their initial conditions (while keeping the initial radial velocity equal to zero).  Since the system is spherically symmetric, the orbits will be restricted to a plane and thus the motion will be 2D.  Figure \ref{fig:spherical_2d} shows characteristic orbits that arise under these conditions.  The orbits were integrated in time scales large enough to clarify their main properties.  We find the following classes of orbits:
\begin{enumerate}[(a)]
\item \textit{Inner orbits (left panel, blue curve):}  in the region that lies inside the first zero of the force profile, the dynamics is as in any other theory that has attractive gravitational force.  The orbital structure includes all the richness that exists in standard gravity, including box orbits, a large number of resonances, loops or rosettes depending on the initial conditions provided.

A possible galactic model taking into account only these orbits can be created by assuming a value for the mass parameter $\mu$ that is small enough for the field oscillations to move to very large radii, outside the galaxy.  Since the entire galaxy will be contained in this region, its dynamics will be very similar to the one we encounter in the standard MOND theories.  An alternative model is discussed in Section \ref{section:alternative}.
\item \textit{Unstable orbits (left panel, orange curve):}  This kind of orbits is located in the empty region of Figure \ref{fig:orbits}.  They are unstable no matter the tangential velocity that may have and will irremediably move away from the galaxy.
\item \label{low_l_loop} \textit{Low angular momentum ring loops (central panel, blue curve):}  
Particles with initial conditions given in the outer part of stable rings (i.e. in the region of the rings where the force is attractive) and have a small initial tangential velocity $v_y$ will fall towards the galaxy and have radial oscillations whose period is much smaller than the orbital period.  These orbits will cross the minimum of the potential and will visit the inner part of the ring where the force is repulsive.  Seen from the center of the galaxy, these orbits will follow an overall circular path, but at the same time, they will have a strong radial component.  From an observational perspective, these may be confused with box orbits that have incursions towards the center of the galaxy.  However, these orbits always remain at a radius confined by the potential peaks, oscillating around the local minima of the potential. 
\item \textit{High angular momentum ring loops (central panel, orange curve):}  For particles that lie in the outer region of the stable rings, the force is attractive and thus it is possible to define a circular velocity. Increasing the tangential velocity above this value (but below the escape velocity) gives rise to orbits whose radial frequency is comparable to the orbital frequency.  These will be similar to standard loops with radial variations that will never be larger than the thickness of the ring. 

Note that owing to the presence of repulsive forces, these orbits cannot exist in the inner part of the stable rings (i.e. the region that lies between a maximum and a minimum of the potential).  In this region, it is only possible to obtain low angular momentum orbits as described in previous point.

\item \textit{Resonances 1:$n$ (two orbits in the right panel):}  We show this class of resonances, not because these are the only existing resonances, but to give a flavor of the possibilities of the theory. The notation 1:$n$ makes reference to $n$ radial oscillations for each angular (or orbital) period.

In order to clarify the distribution of resonances in velocity space, we integrated a large number of orbits varying the initial tangential velocity ($v_y$) while keeping the initial value of the radial velocity ($v_x$) equal to zero.  For an orbit to be in any of these resonances it has to arrive to its point of departure with the same velocity.  We show the result of these integrations in Figure \ref{fig:spherical_2d_resonances}.  The blue curves correspond to the difference between the initial and final position after one orbital period; the orange curves are the final radial velocity in the $x$ direction (note that the orbits were launched from the $x$ axis, so at that point the components $x$ and $r$ have the same values).  The velocities that correspond to the resonances 1:$n$ are the places where both curves are zero (we highlighted them with vertical thin lines).  We identified the $n$ value of each resonance by dividing the total angular period obtained from the numerical solutions by the time difference between the starting time and that of the first maximum of the radial component of the position (these are easy to estimate given the numerical solution).  In order calculate the orbital period we employed the \texttt{solve\_ivp} routine of \texttt{scipy} for the integration of the orbits, which can apply a root finding algorithm to identify specific points in the solution (in this case, the point in which the position crosses the positive part of the horizontal axis in the vertical direction).

Since we are providing initial conditions in the outer part of the ring, it is possible to calculate the circular velocity as it is usually done in any central force system:  $v_c = \sqrt{r \mathrm{d}\Phi/\mathrm{d}r}$.  The resulting values are not necessarily related to the resonances and coincides with the place where $x-x_0$ changes its sign (see thick vertical line in Figure \ref{fig:spherical_2d_resonances}).  The circular velocity decreases when we move the initial condition in configuration space towards the minimum of the potential, where $\nabla\Phi = 0$.  For orbits that start near the minimum, the amplitude of the radial oscillations of the 1:$n$ resonances decreases with $n$.  The circular orbit associated to the minimum corresponds to the limit of these resonances when $n \rightarrow \infty$.

Finally, note that these resonances 1:$n$ can also be divided between high and low angular momentum orbits.  The first will be similar to the typical loop orbits, while the others will have a large number of radial oscillations in each angular period.
\end{enumerate}
The classification of 2D orbits provided in this section corresponds to the central region of the potential, and the innermost unstable and stable rings (i.e. up to the end of the first green region from left to right in the third panel of Figure \ref{fig:orbits}).  The same orbital structure repeats itself as we move to the right towards the outer rings.

\begin{figure*}
    \includegraphics[width=\textwidth]{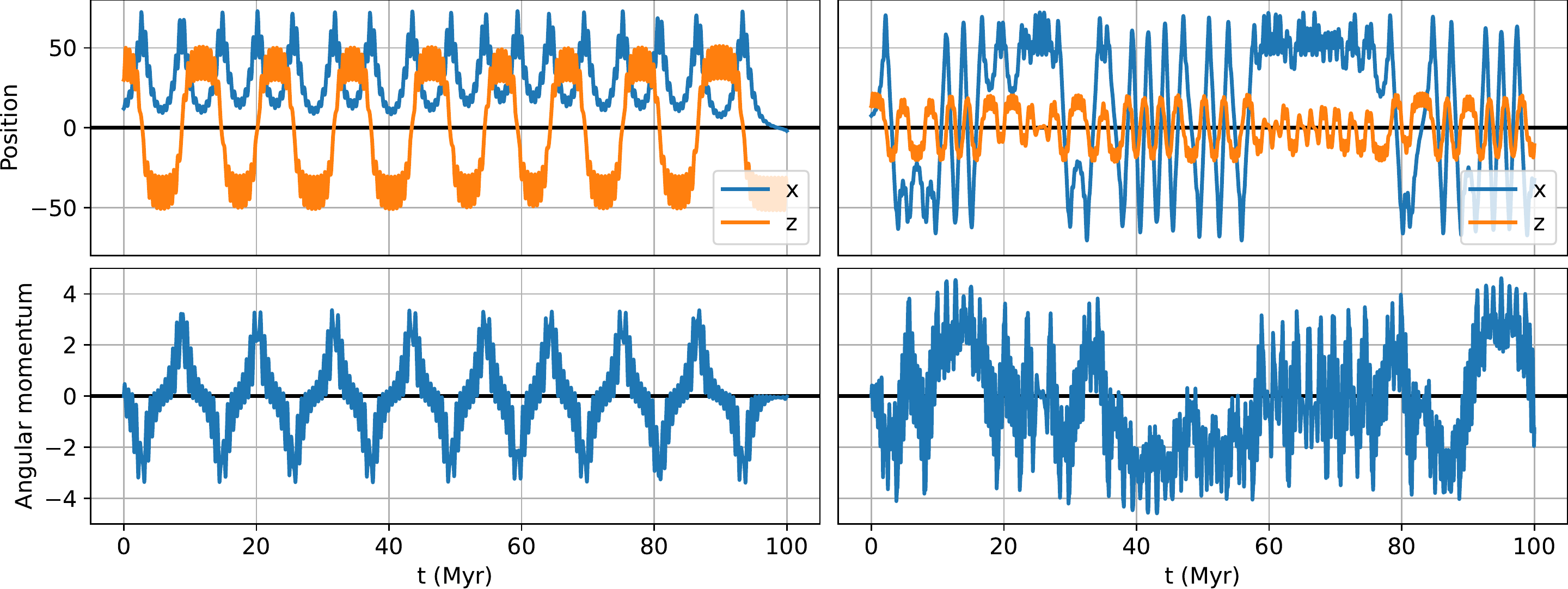}
    \caption{Time evolution of coordinates (upper panels) and angular momentum in the $z$ direction (bottom panels) for the regular (left) and chaotic (right) ring orbits.  The initial conditions are the same that produce orange orbits in the central and right panels in Figure \ref{fig:flattened_2d_box}, although the integration time is $10^5$ Myr.}
    \label{fig:flattened_2d_time}
\end{figure*}

\subsection{2D orbits in flattened potentials}

An additional step in complexity can be reached by flattening the system studied in previous section (while staying in a 2D plane that contains the preferred axis).  The orbits in {\MG} follow the same behavior as in standard gravity in the sense that they can be divided in two large families of box and loop orbits.  We describe these two families separately and highlight particulars of these orbits when they are located in the rings that arise in the potential when the mass parameter $\mu$ is different than zero.  The flattening of the systems is defined as described by Eq.~\ref{def_flattened_force}.  During this entire section we assume $(b,c)=(1,0.7)$ unless a different value is mentioned in the text.

As a reminder to the reader, we define \textit{box} orbits as these orbits for which the angular component of the position of the test particles oscillates between two finite values moving back and forth.  This change in direction results in a change of the sign of the angular momentum, which has a mean value in time equal to zero (when looking at the system in a non-rotating frame).  These orbits exist in opposition to the \textit{loop} orbits in which the angular component of the position evolves monotonically.  In these cases the test particles circulate in a fixed direction around the potential center.

\begin{figure*}
    \includegraphics[width=\textwidth]{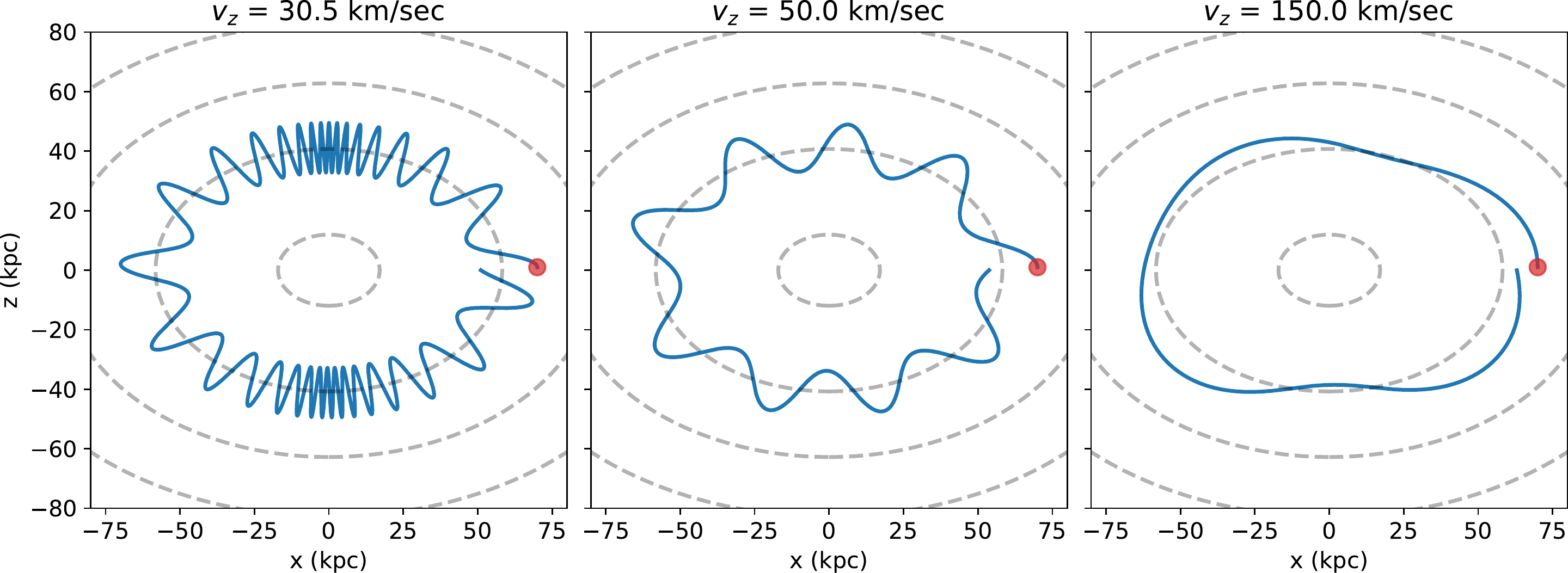}
    \caption{Examples of loop orbits in the stable rings of a flattened 2D system.  The Figure is equivalent to Figure \ref{fig:flattened_2d_box}, but here we provide the particle with an initial velocity in the positive $z$ direction (the panels differ on the initial velocity, which is shown above them in km/sec).  The orbits are integrated during an entire orbital period.  The ellipsoid is oriented with its minor axis in the $z$ direction.  We show orbits in the $x-z$ plane.  The dashed lines are the zeros of the force.}
    \label{fig:flattened_2d_loops}
\end{figure*}

\subsubsection{Box orbits}

Figure \ref{fig:flattened_2d_box} shows examples of box orbits in {\MG} for different initial conditions in configuration space (the initial velocity is zero in all these cases).  The left panel contains orbits that start near the equator of the system.  The inner orbit (which is located in the region inside the first minimum of the potential) is a typical banana orbit which is known to exist in standard gravity.  For outer initial conditions that start near the second minimum of the potential (orange curve in the left panel), the orbit has a box behavior in the sense that the angle that corresponds to a description of the orbit in polar coordinates does not cover the entire available domain from 0 to $2\pi$.  Thus, the particle's position not only oscillates in radius as in previously studied orbits, but also in latitude, changing direction periodically.  Note that the orbit is very similar to orbits studied 2D systems in standard gravity in the plane $R-z$, however, here we are showing Cartesian coordinates $x-z$ and assume $y=0$.

The central panel in the same figure shows two orbits that start at higher latitudes.  Again, the inner orbit is a classical box orbit.  Since in this region the {\MG} potential mimics the MOND potential (which in turns mimics the Newtonian potential when a dark matter halo is added to the system), varying the initial condition in this region will show the existence of numerous resonances and chaos as in the standard gravity case.  The outer orbit follows the same behavior as in the left.  However, there is a clear variation of the radial oscillation frequency when interpreting it as a function of latitude.  Note that the variation of the radial oscillations with latitude does not necessarily imply a variation of the frequency of the radial oscillations with respect to time.

Increasing the latitude of the initial conditions even more brings the orbit into a chaotic region of phase-space.  We show an example in the right panel of Figure \ref{fig:flattened_2d_box}, where we increased the flattening of the system, so we can show the effect with small integration times (we used $(b,c)=(1,0.3)$ for this particular orbit).  The presence of chaos in this region of phase-space provides the system with a Lorenz attractor behavior:  the orbit switches at random times between two states defined by the orbits described in previous panel.  The attractive points will be in this case the two minima of the potential that are located on the $x$ axis.  The transfer between box orbits with positive or negative values of $x$ is made through the ellipse associated to the minimum of the potential that defines the ring.  In this sense, we can say that due to chaos, orbits switch between loop orbits that follow the minimum of the potential and boxes with positive or negative values of $x$.

A clearer understanding of how these chaotic orbits evolve can be obtained from Figure \ref{fig:flattened_2d_time}.  The upper panels show the time evolution of the Cartesian coordinates $x$ and $z$ for the same ring orbits shown in the central and right panel of Figure \ref{fig:flattened_2d_box}. The integration time however is much larger than what is shown in Figure \ref{fig:flattened_2d_box}. The bottom panels show the evolution of the $y$ component of the angular momentum $L$ (i.e. the component perpendicular to the plane that contains the orbit).  We can see that in the case of the regular orbit (left panels), the $x$ coordinate is always positive while the $z$ coordinate changes sign in each orbital period.  These slow oscillations in $z$ are perturbed with the high frequency oscillations of the radial component around the minimum of the potential.  The angular momentum is not conserved and oscillates between positive and negative values according to the direction of the orbit.

The right panels of Figure \ref{fig:flattened_2d_time} show the evolution of all these quantities in the chaotic case.  There is no regular pattern and the orbit switches at random times between loops and boxes.  For instance, in the interval that approximately goes from 40 to 60 Gyr, the $x$ component of the position changes sign periodically while the sign of the angular momentum is fixed (this shows that at these times, the orbit is behaving as a loop).  The situation changes in the interval that goes from 60 to 80 Gyr:  the sign of $x$ is always positive and the angular momentum switches sign each time the evolution of the angular component of the position changes sign (i.e. the orbit behaves as a box).  At other times, the orbit changes randomly between these two states.

\begin{figure*}
    \includegraphics[width=\textwidth]{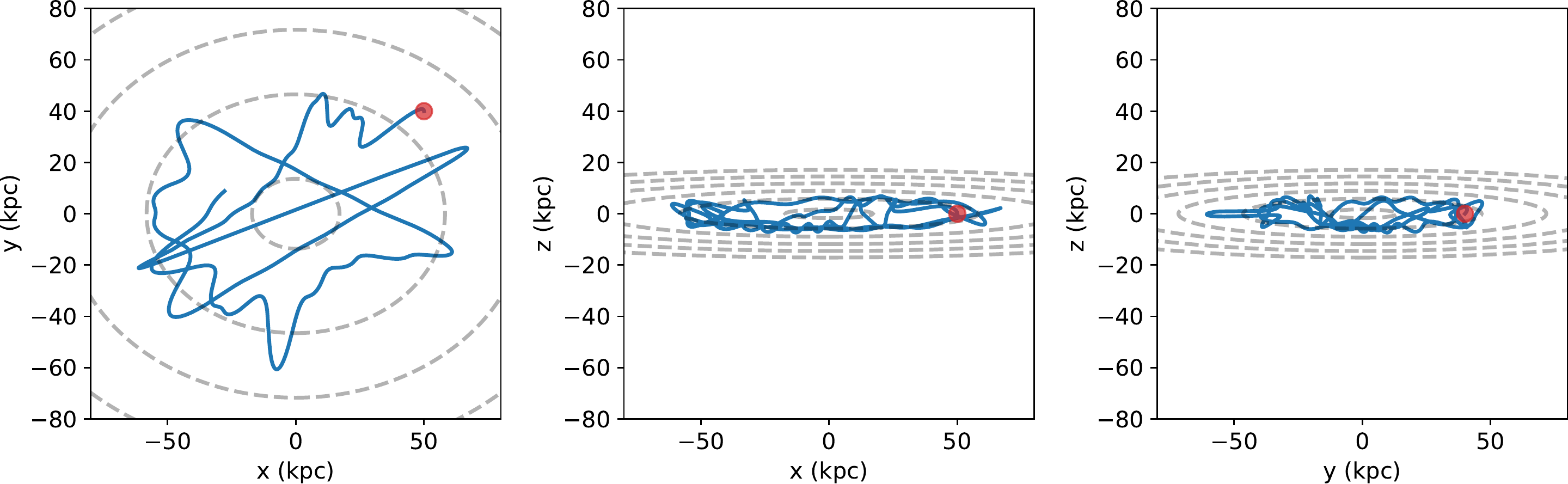}
    \caption{Examples of a 3D chaotic orbit in a perturbed disk.  Each panel shows a different projection of the same orbit.  
    The dashed lines at the minima of the potential (i.e. the zeros of the force distribution).  For ease of readability we show only a few of these zeros in the lateral projections. Note that the scale is the same in all the axis.}
    \label{fig:disk_3d_chaotic}
\end{figure*}

\subsubsection{Loop orbits}

We now switch back to regular (i.e. non-chaotic) orbits, in this case loops.  Figure \ref{fig:flattened_2d_loops} shows a few examples of these orbits, which, for simplicity, were integrated during only one orbital period.  In the inner region these orbits work as in standard gravity.  When moving outwards into the rings, we find that for low angular momentum (left panel), the radial oscillations are very similar to the case of box orbits: the angular frequency of the radial oscillations increase as a function of the latitude ($z$-axis).  The difference now is that there is enough angular momentum for the particle to cross from positive to negative values of $x$ and thus, the entire loop can be formed.  Note that the initial and final position in this example is not the same (in other words, radial and angular oscillations are not commensurable).  A priori one could expect that the entire ring will be eventually covered by the orbit while integrating up to $t=\infty$, however, resonances do exist in this family of orbits.  Thus, there is no warranty that this will happen.

The central panel of the same figure shows an example with larger angular momentum.  In this case, the dependence of the radial frequency with latitude becomes less obvious and the orbit behaves as a loop orbit in a spherical system that follows the minimum of the potential (see for instance the blue orbit in the central panel of Figure \ref{fig:spherical_2d}).  For even larger values of the angular momentum (right panel), the period of the radial oscillations can be comparable to the period of the angular component.  

\subsection{3D orbits}
\label{section:3d_orbits}

The complexity of the orbital structure in 3D (specially in the case of triaxial potentials) is such that deserves a targeted study that is well beyond the limitations of this short paper.  In order to give a flavor of how these orbits behave, we present an example of a chaotic orbit in a perturbed (i.e. not symmetric) disk.  We approximate the potential of the disk by flattening the same gravitational potential that we discussed in previous sections.  The ellipticity are $(b,c) = (0.8, 0.1)$, which means a very strong flattening in the $z$ direction (i.e. the system is a disk) and a mild flattening in the $x-y$ plane.  Figure \ref{fig:disk_3d_chaotic} shows three projections of this particular orbit.  The initial conditions in configuration space are highlighted with a red dot.  As for the velocities, we gave equal kicks in the $y$ and $z$ direction of 33.2 km/sec.  The orbit is clearly chaotic.  We can see that in $x-y$ projection, the orbit tries to follow the ellipse that corresponds to the minimum of the potential as it happens in 2D orbits, but from time to time jumps in a random direction towards a different location of the same ellipse.  The transfer is made following the zeros of the potential in the other projections and thus, the orbit does not cross the center of the galaxy, but moves above and below it.

\begin{figure}
    %REFEREE
    %\includegraphics[width=0.8\columnwidth]{orbits_incoming.pdf}
    \includegraphics[width=\columnwidth]{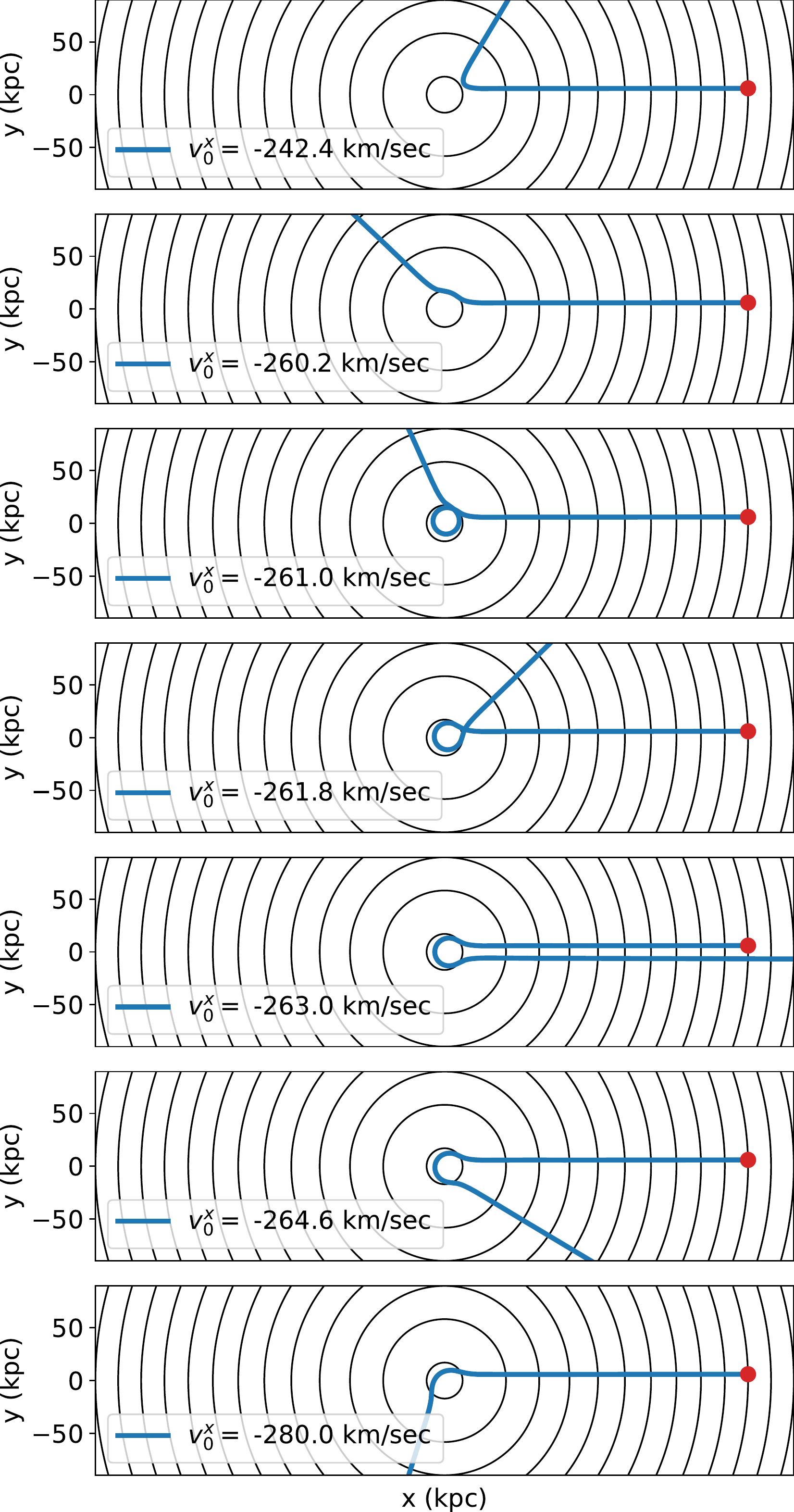}
    \caption{Examples of incoming orbits and various types of deflection around a spherical galaxy.  The initial position of the incoming particle is $x=288$ kpc and $y=6$ kpc, so the orbits are not strictly radial.  The panels differ on the initial velocity in the $x$ direction.  The black circles correspond to the zeros of the force profile.}
    \label{fig:incoming}
\end{figure}

\subsection{Incoming orbits}
\label{section:incoming}

Previous sections discuss orbits associated to galactic systems or to the systems of shells that may surround them in a {\MG} universe.  Here we discuss incoming orbits that would be followed by material as it is accreted by galaxies or clusters of galaxies in a hierarchical structure formation scenario.  For a particle to be accreted by a system, two things must happen: firstly, the particles must be able to approach the system.  Secondly, said particles must be able to stay in the system after arrival.  In the standard gravity case, this process depends on the initial velocity of the particles.  Since the force is always attractive, all particles with an initial negative velocity will be able to approach their potential host systems.  However, only particles whose velocity is smaller that the escape velocity of the systems will be effectively accreted.  This can be clearly seen in the upper panel of Figure \ref{fig:orbits}: only particles approaching through the blue region of the panel will stay in the galaxy after the interaction occurred.

In the standard MOND case (central panel of Figure \ref{fig:orbits}), such distinction does not exist.  The presence of a logarithmic potential, will force all particles to be bound to any isolated system. The third panel of the same figure shows that the situation is much more complex in {\MG} even for the simplest case of a particle approaching a spherical system in purely radial orbits.  Unlike the other two gravitational models, the first condition for accretion (i.e. ability to approach a system) depends on the initial velocity.  For initial velocities such that the particles are located in the green blobs shown in Figure \ref{fig:orbits}, the radial component of the trajectories will oscillate and thus, no approach is possible.  It is still possible to approach the systems through the empty or orange regions.  However, in none of these cases, an incoming particle will be able to stay in the central region of the potential (blue region in the figure).  In the first case (empty regions), the particles will approach until a minimum radius is reached and then will be expelled.  In the other case, the particle's velocity is larger than the escape velocity and thus, the particles will enter the systems, but will pass through them and continue their trip up to infinite.

Since it is not possible to approach a system through a stable orbit in the purely symmetric case, the symmetry must be broken somehow if structure formation in {\MG} is going to be hierarchical.  Here we study the simplest case that consists in substituting radial orbits with orbits that have an impact parameter different from zero.  We present in Figure \ref{fig:incoming} a series of incoming trajectories for the same spherical profile described in Section \ref{section:orbits}.  Each panel shows trajectories that approach from $x\sim 288$ kpc with an impact parameter of 6 kpc (i.e. the particle does not approach in a purely radial orbit, but has an initial trajectory that is shifted in the vertical direction).  Different panels correspond to different initial velocities in the horizontal direction (the initial vertical velocity is zero in all cases).  We find the following types of incoming orbits:

\begin{enumerate}
\item \textit{Reflected trajectories (first panel):}  The upper panel of Figure \ref{fig:incoming} shows that the reflection we found in the case of purely radial orbits approaching through the empty regions of Figure \ref{fig:orbits} can also occur in the case where the impact parameter is different from zero. The bounce will occur at different shell depending on both the initial incoming velocity and the geometry of the collision (i.e. the impact parameter).
\item \textit{Ejection after interaction with the boundary of the innermost region (second panel):} In this case, the velocity is such that the particle can approach the innermost equilibrium point (which is unstable and, depending on the galactic model assumed, may define the boundary of the galaxy).  Once the particle has reached the sphere that corresponds to this point, it can move tangentially in an unstable region until is eventually ejected.
\item \textit{Unstable loops in the inner region (third and fourth panels):}  For larger velocities, the particle can actually enter the innermost region (where dynamics are similar to that of classical MOND). However, these velocities are too large for the particles to stay in the system, and thus, they are ejected after orbiting the center of the galaxy once.  Note that the final trajectory is similar to those shown in previous panels, but here the particle has had a chance to interact with the galaxy.  
\item \textit{Reflection in the inner region (fifth panel):} The empty regions of Figure \ref{fig:orbits} that we already discussed at length correspond to purely radial orbits that are reflected before they arrive to the central system.  The incoming orbit that we included in this panel shows that such reflection can also occur for non-radial orbits.  However, in this case, the reflection does not occur before the central system is reached, but after passing behind it.
\item \textit{Standard collision in an attractive field (last two panels):} For larger velocities, the regions of repulsive force become less important and the particle behaves as in a textbook collision with an attractive force. The trajectory is bent, but no other major perturbations exist (although, note that due to the peculiar form of the potential, we do not recover the usual hyperbolic orbits).  Increasing the velocity further, will result in trajectories with smaller and smaller deflection angles.  In the limit where $v_0 \rightarrow \infty$ where the particle will cross the galaxy following a straight line without being perturbed at all.
\end{enumerate}

These orbits show that it is possible to approach spherical systems that were a priori protected by approaching them in a non-radial orbits.  None of the orbits we show in these experiments stay inside the central system.  However, we need to take into account that dynamical friction was not included in the calculations.  This effect may play a fundamental role when in comes to allowing incoming substructure to be accreted  by their hosts.

\section{Discussion}
\label{section:discussion}

\subsection{Implications for non-linear cosmology}

One of the challenges (puzzles in the terminology employed by \cite{kuhn1970}) that the classical MOND family of theories is facing is that of the presence of oversized structures in the context of non-linear structure formation in cosmology.  Several authors have shown using non-linear cosmological simulations that the normalization of the non-linear power spectrum of density perturbations $\sigma_8$ at redshift $z=0$ is larger than the observed value by a factor that can be larger than two and well outside the observational error bars \citep{Nusser02, llinares_thesis, 2016MNRAS.460.2571C}.  A very likely reason for this phenomenon is that the MONDian force decreases too slowly when moving far away from overdense regions.  This means that the gravitational force felt by the material that is located inside voids is too large and so it forces matter to be quickly expelled from the voids and accumulate in high density regions at a rate that is much larger than the one required by observations.  This effect is related to the fact that extrapolating the requirement of flat rotation curves to infinitely small accelerations results in a gravitational potential that is logarithmic all the way to infinity, and thus, results in an infinite escape velocity from any isolated potential well.

A phenomenological solution proposed by \cite{llinares_thesis}, consists in redefining the asymptotic behavior of the interpolation function $f$ that defines the transition between standard physics and the MOND phenomenology (see Eq.~\ref{eq_2_mond}).  The new proposed function has a standard MOND behavior in the observed domain of accelerations (i.e. Newtonian and MOND regimes at high and low accelerations respectively) and an additional Newtonian regime at extremely low accelerations (below the limit reached by the smallest galaxies, which is of the order of $10^{-2} a_0$).  Under this new interpolation function, the MOND effect works as usual inside galaxies (and thus, can provide them with flat rotation curves) and has a cut-off far away from galaxies and into the regions where the MONDian void problem exists.  Preliminary simulations show that this solution can save MOND from the void problem \citep[see Appendix A in][]{llinares_thesis}. The question we would like to investigate here is if any of the effects described in previous sections can provide alternative solutions to this problem whilst maintaining the original definition for the interpolation function.

The presence of repulsive forces outside galaxies in {\MG} may indeed provide an alternative solution to this problem.  In this case, the screening of the MOND force that may be required at very low accelerations is not encoded in the shape of the free function $f$ as proposed by \cite{llinares_thesis}, but in the oscillations induced by the mass term in the field equations (Eqs.~\ref{eq_1_mond} and \ref{eq_2_mond}).  Thanks to these oscillations, the mean force that a particle feels when approaching a galaxy is reduced with respect to standard MOND.  

While these shells of repulsive force can help in solving the MONDian void problem, they may create a different problem.  We provided examples in Section \ref{section:incoming} that show that the additional velocity that may be required to penetrate these shells and access galaxies may be too large and thus, particles may be ejected after reaching the inner regions of the potential.  This may imply that galaxies in {\MG} become isolated from the rest of the Universe and that no accretion of material is possible after the first collapse occurs.  This may be in contradiction with the usually assumed hierarchical model of galaxy formation in which galaxies are formed by adding up material from incoming substructures.  However, we need to take into account the following facts:

\bi
\item Here we studied the trajectory of incoming individual particles and did not take into account their own potential.  In the real case, the incoming object will not be a point particle, but an extended massive object, which will give rise to an additional extended potential.  The interaction between this potential and that of the main object is non-linear and may perturb the original potential in such a way that the repulsive forces will disappear and allow low velocity material to be accreted by the more massive object.
\item We are assuming that the accretion of material occurs isotropically and in isolation (i.e. the interaction occurs only between the massive object and the incoming particles).  In the real case, the accretion occurs not only from the voids, but also through filaments which have their own gravitational potential.  These filaments may create channels where material can be accreted, while the other directions will be protected by the presence of repulsive forces.  This may solve the MONDian void problem while keeping the hierarchical mechanism in place.
\item Panels 3 to 7 in Figure \ref{section:incoming} show that it is possible to reach the galaxy in spite of the presence of rings where the gravitational force is repulsive.  Dynamical friction may contain these unstable orbits and allow incoming particles to stay inside the galaxies.  This effect has been studied in classical MOND and was found to be less effective that in Newtonian gravity \citep{2008MNRAS.386.2194N}.  The question of the validity of these results in {\MG} is still open.
\ei

Calculating the actual impact of these solutions in the non-linear matter power spectrum in cosmology will require running fully non-linear cosmological simulations.  This is a challenging enterprise which requires not only dealing with a non-linear version of Poisson's equation (as in standard MOND), but also to include oscillations in the fields, which is well beyond the scope of this paper.

\begin{figure}
    \includegraphics[width=\columnwidth]{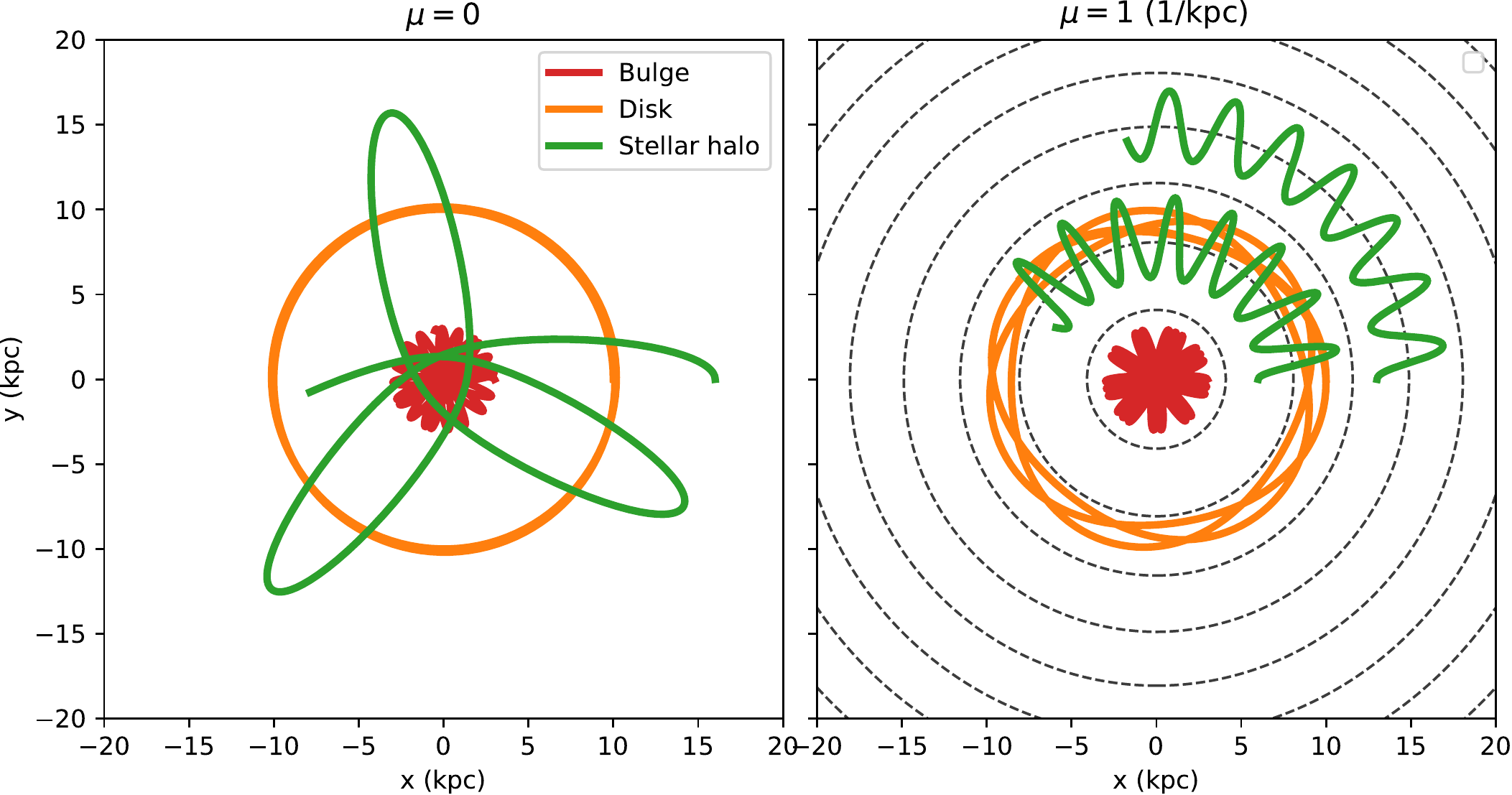}
    \caption{Left: structure of a disk galaxy assuming $\mu=0$, which means that the model has a phenomenology equivalent to MOND (which is similar to the Newtonian case after a dark matter halo is added).  We provide examples of bulge, disk and stellar halo orbits.  Right, alternative proposal that arises when assuming a large value for the mass parameter (we used $\mu=1 $ kpc$^{-1}$).  See section \ref{section:alternative} for details.}
    \label{fig:new_model}
\end{figure}

\subsection{An alternative view of galactic structure}
\label{section:alternative}

In previous sections we assumed that galaxies are located in the innermost stable regions of the potential profiles (i.e. blue region in the third panel of Figure \ref{fig:orbits} or \ref{fig:spherical_2d}).  The repulsive force ring that surrounds these regions (empty regions in same figures) provides a natural mechanism to truncate galaxies or their effective dark matter halos.  However,  constraints on the mass parameter $\mu$ are not yet developed, so we do not know what is the real impact that these repulsive forces will have on the dynamics of the systems and on their observable properties.  Here we speculate on an alternative scenario which arises when we assume a value of $\mu$ that is large enough for the rings described in previous sections to be located inside the galaxies.

In order to fix ideas, we show in the left panel of Figure \ref{fig:new_model} the standard view of a spiral galaxy, which is based of the following three structures: one or multiple disks with rather circular orbits in a plane (orange), a bulge (red) with a large fraction of radial (i.e. box) orbits akin an elliptical galaxy, and a stellar halo (green) with the same characteristics, but orders of magnitude larger.  From the point of view of the actually observations, we need to take into account that we cannot measure individual orbits of stars over time, but only instantaneous quantities such as velocity dispersion profiles or rotation curves, both of which combine information associated to a large number of stars.  A priori, there is no reason to assume that what we see as orbits pointing towards the center of the galaxy (as in the case of the green orbit in the left panel of Figure \ref{fig:new_model}) are actually orbits that do in fact reach the center.  Alternatives could exist and in fact, the theory we study here provide ingredients for building one of them. 

Such alternative galactic model can be obtained by increasing the mass parameter $\mu$ to the point in which the first unstable point is located, for instance, in the boundary of the bulge of the galaxy.  The right panel of figure \ref{fig:new_model} shows a possible decomposition of a spiral galaxy in this framework.  The bulge (not the entire galaxy) will be located in the innermost region of the potential and thus, will be a standard bulge.  The disk orbits will be the large angular momentum ring circular orbits and resonances with low $n$ shown in Figure \ref{fig:spherical_2d} (see orange orbits in the central and right panels).  Finally, what we interpret observationally as radial stellar halo orbits will be perturbed circular orbits living in the stable rings and which have fast oscillations in radius (see blue orbits in central and right panels of Figure \ref{fig:spherical_2d}, orange orbits in Figure \ref{fig:flattened_2d_box} or orbits shown in the left and central panel of Figure \ref{fig:flattened_2d_loops}).  The important difference is that these halo stars will not constantly migrate through the galaxy towards the center, but will have a fixed radius with fast, but small perturbations.  So what from an observational perspective may appear to have fast movement towards the center of the galaxy, may actually be an oscillating circular orbit.

Confirmation of the model and eventual determination of observable quantities that could be used to compare it with the standard model will require detailed  Schwarschild modeling \citep{1979ApJ...232..236S} or N-body simulations.  Note that the Schwarschild technique was confirmed to be valid for the standard MOND theory in \citep{2009MNRAS.396..109W}, however, it was never implemented with repulsive forces.  

\subsection{Observational consequences}

We have shown in previous sections that oscillations that arise in the force distribution thanks to the mass terms in the field equations result in repulsive forces and unstable regions in phase space where particles may be expelled from galaxies.  This force distribution may result in galaxies being surrounded by a combination of gaps and overdense regions where particles are trapped in perturbed circular orbits (which may have strong radial oscillations).  Depending on the value of the mass parameter $\mu$, the entire pattern can stay outside the galaxy (for small $\mu$) and thus, have no more impact than a tinny perturbation in lensing observables.  Alternatively, large values of $\mu$ may bring the oscillations inside the galaxy.  The absence of observed gaps in galaxies may be used as an argument to constrain $\mu$ however, careful modeling must be done before excluding these regions of the parameter space.

We expect that the accumulation of material in the galactic rings will be minimal as an accumulation of mass in this region would in itself perturb the ring structure. Further work is needed to determine the limit at which the accumulated mass and thus, the perturbations collapse or alter the spacial oscillations in the potential in such a way as to either eject the material or allow the material to migrate into the host galaxy. The use of photometric arrays such as the Dragonfly Telephoto Array \citep{2014PASP..126...55A} or the Huntsman Eye Telescope \citep{2019arXiv191111579S} may afford an opportunity to detect what are expected to be extremely low surface brightness features.

Regarding the 3D disk orbits, we showed in Section \ref{section:3d_orbits} that there is a family of orbits that moves in radial directions, not inside the disk itself, but above or below it (see Figure \ref{fig:disk_3d_chaotic}).  Thus, the theory predicts that, in a particular region of its parameter space, material (stars and gas) should be moving following flattened stable rings that may exist above and below disks.

\section{Conclusions}
\label{section:conclusions}

We present a phenomenological study of allowed stellar orbits in the latest relativistic extension of the MOND paradigm, which we refer to as {\MG} \citep{Skordis_Zlosnik_2021, Milgrom83, 1984ApJ...286....7B}.  This particular theory is special thanks to its ability to provide for the first time a good fit to background cosmology, CMB observations and the matter power spectrum in cosmology.  It is also special because its field equations include a mass term, which largely enriches the phase-space structure of free particles.

Thanks to the mass term, the gravitational potential (and hence the force that is responsible for particle trajectories) develops spacial oscillations with additional minima \citep{Verwayen}.  We find that these new potential wells give rise to a new family of orbits whose radius can oscillate around the minima of the potential without having to necessarily migrate all the way to the center of the galaxy.  We describe the main effect in the simplest possible potential (a 1D spherical galaxy) and increase complexity by adding new spacial components and flattening in one or two directions.

Far from being a mathematical curiosity these results should have a fundamental impact in various aspects of MOND.  For instance, in the context of non-linear cosmological evolution, we know that a simple extrapolation of the classical MOND theory towards cosmology fails to reproduce the observed galaxy power spectrum at redshift zero.  This is because the logarithmic potential that defines MOND results in the gravitational force being far too large in voids, and thus, structure formation that is too violent when compared with the observable Universe.  A typical cosmological simulation with MOND does not predict a cosmic web as observed, but enormous voids surrounded by very few and very large objects.  The spacial force oscillations that we find in R-MOND may help to effectively shutdown the logarithmic force inside voids and thus, alleviate the problem.

In order to test this hypothesis, we calculated incoming orbits with several initial velocities towards a galaxy.  We find that indeed, the galaxy is protected against incoming material.  In fact, we find that it may be too protected and thus, accretion may be impossible once galaxies form.  According to the exact value of the mass parameter $\mu$, the theory can lead to two distinct scenarios:  excessive accretion leading to the MONDian void problem or no accretion at all, which will force us to rely on monolithic structure formation models.  Finding out if it is possible to have an intermediate model that is compatible with observations will require running complex non-linear structure formation simulations.  Thanks to the presence of the mass term, solvers that are already able to deal with the complex non-linearities that exist in classical MOND \citep[e.g.][]{Nusser02, Knebe04b, Llinares:2008ce, llinares_thesis, 2015CaJPh..93..232L, 2016MNRAS.460.2571C} will have to be re-written from scratch.

For values of $\mu$ that can have an impact on the solutions, we find two additional regimes.  For low values of $\mu$, galaxies behave as in standard MOND, but with the logarithmic behavior of the potential outside the galaxies substituted with oscillations.  In the other extreme, at high values of $\mu$, the oscillations can occur inside the galaxies, and effectively quantize the radial orbits.  Confirmation of what situation is more compatible with observations will require the application of complex N-body or Schwarschild techniques to develop self-consistent galaxy models.

The phenomenological results presented in this paper show that R-MOND is not simply a formal extension of MOND to force it to fulfill covariant requirements, but it is a distinct theory which produces distinct predictions.  Depending on what exact value that may have been chosen by nature for the mass parameter of the theory, the new phenomenology that we found may give us a chance to find undisputable evidence for the presence of something beyond General Relativity in the data.

\begin{acknowledgements}
    C. Llinares acknowledges support from the Funda\c{c}\~{a}o para a Ci\^{e}ncia e a Tecnologia (FCT) through the Investigador FCT Contract No. CEECIND/04462/2017 and POCH/FSE (EC).  Llinares thanks Prof. Daniel Carpintero and Peter Verwayen for relevant comments that greatly improved the quality of the article.
\end{acknowledgements}

\bibliographystyle{aa} % style aa.bst
\bibliography{../references.bib} % if your bibtex file is called example.bib

\end{document}